# *A compact 3.5-dB squeezed light source with atomic ensembles*

GUZHI BAO,[1] XIAOTIAN FENG,[1] BING CHEN,[1] JINXIAN GUO,[1] HENG SHEN,[1*] LIQING CHEN[1*], AND WEIPING ZHANG[1]

[1]*Quantum Institute of Atom and Light, State Key Laboratory of Precision Spectroscopy, Department of Physics, East China Normal University, Shanghai 200062, P. R. China.*

*\*hengshen@nbi.dk*

*\*lqchen@phy.ecnu.edu.cn*

**We reported a compact squeezed light source consisting of an diode laser near resonant on 87Rb optical D1 transition and an warm Rubidium vapor cell. The -4±0.2 dB vacuum squeezing at 795 nm via nonlinear magneto-optical rotation was observed when applying the magnetic field orthogonal to the propagation direction of the light beam. This compact squeezed light source can be potentially utilized in the quantum information protocols such as quantum repeater and memory, and quantum metrology such as** *atomic magnetometer.*

Surpassing the standard quantum limit in the measurement has always been a long-term goal for quantum metrology and quantum communication. Squeezed states of optical fields are important resources for photonic quantum technology particularly with continuous variables [1-4]. Recently, squeezed vacuum states of light were used to improve the sensitivity of a gravitational-wave observatory in GEO600 project [5]. So the squeezed vacuum lights have recently been the subject of some attention. The basic mechanism for vacuum squeezing in a single pass linearly polarized pump configuration is polarization self-rotation (PSR)[6] which can be explained by solving Heisenberg-Langevin equation describing the propagation of quantized field through atomic ensemble[7]. Until today, squeezed vacuum generated in a rubidium vapor has been limited to under-2.9dB[8]. In this letter, we report ±4_0.2 dB vacuum squeezing at 795 nm via nonlinear magneto-optical effect in Rubidium vapors. Compared to traditional optical parametric oscillator (OPO) [1], the experimental setup for this kinds of vacuum squeezing is simpler and more compact, where only a warm vapor cell and one laser beam near resonant on D1 transition are utilized. Additionally, since the squeezed light is generated in the atomic ensemble, this nonclassical light can be directly applied in the quantum protocols where atomic ensembles are involved such as quantum memory [9] and atomic magnetometer[4].

The configuration of the squeezed light source considered here is based on the nonlinear magneto-optical rotation where the polarization direction of an input field is rotated due to the resonant light-induced alignment of atomic spins. Such interaction leaves the quantum fluctuations of the optical field unchanged, however, squeezed vacuum state of optical field is produced in orthogonal polarization [6, 8, 10, 12, 14, 15], which is known as polarization self-rotation squeezing. We consider the 87Rb atoms interact with a field resonant on atomic transitions. Following the derivation in [7], the optical field is described by slowly varying canonical operators with the annihilation (creation) operators being denoted $\hat{a}_\pi(\hat{a}_\pi^+)$. And one can define continuous atomic operators at $\hat{\sigma}_{\mu\nu}(z,t)$ position z by averaging over a slice of the atomic medium of length $\triangle z$ as $\hat{\sigma}_{\mu\nu}(z,t) = \lim_{\Delta z \to 0} \frac{L}{N\Delta z} \sum_{z \ll z_j \ll z+\Delta z} \hat{\sigma}_{\mu\nu}^j(z,t)$. The Hamiltonian in the interaction picture can be written as:

$$\hat{H} = -\hbar \int \frac{dz}{L} N [\sum_i g_i \, \hat{\sigma}(z,t)\hat{a}_\pi + \sum_j \frac{\Omega_i}{2} \hat{\sigma}(z,t) + \text{H.C.}] \quad (1)$$

with $g_i$ the atom-field coupling constants and $\Omega_i$ the Rabi frequency. The evolution of atomic operators is determined by Heisenberg-Langevin equations [15,16] with the decay rate $\Gamma_{\mu\nu}$,

$$\frac{\partial}{\partial t}\hat{\sigma}_{\mu\nu} = \frac{i}{\hbar}[\hat{H}, \hat{\sigma}_{\mu\nu}] - \Gamma_{\mu\nu}\hat{\sigma}_{\mu\nu} + \hat{F}_{\mu\nu} \quad (2)$$

The Langevin operators $\hat{F}_{\mu\nu}$ are characterized by:

$$< \hat{F}_{\mu\nu}^+(z,t) F_{\mu'\nu'}(z',t') > = 2D_{\mu\nu\mu'\nu'}\delta(t-t')\delta(z-z')$$

$$< F_{\mu\nu}(z,t) > = 0 \quad (3)$$

The field evolution equations are obtained from Maxwells propagation equations under the slowly varying envelope approximation:

$$(\frac{\partial}{\partial t} + c\frac{\partial}{\partial z})\hat{a}_\pi = \frac{i}{\hbar}[\hat{H}, \hat{a}_\pi] \quad (4)$$

The calculation details can be found in [7].

The experimental scheme is presented in Figure 1. The isotopical 87Rb atoms is contained in cylindrical Pyrex cell (75mm in length; 25mm in diameter), which is inclosed in the magnetic shields with 4-layer μ-metal. The output of an external cavity diode laser, whose frequency is near resonant on F = 2 to F'= 2 as the inset of the Figure 1 shown, is coupled into a polarization maintaining single-mode optical fiber and passes through a Glan polarizer to ensure the high-quality of linear polarization in y-direction and single-mode spatial distribution of the laser beam. The collimated laser beam with the waist of 2mm enters into the vapor cell heated at 74°C to generate the signal field, which can be polarization squeezed or vacuum squeezed light field under some certain conditions. In order to quantify the squeezed degree of the signal field, a homodyne detection is built following the vapor cell. The generated signal field polarized in x-direction and the strong drive field with y-polarization can be separated by polarization beam splitter (PBS). And then the drive light is used as a local light to spatially overlap with the weak signal mode. By scanning the relative phase of them with a Piezoelectric Transducer, we can measure the fluctuation of $\hat{X}(\varphi) = \hat{X}cos(\varphi) + \hat{P}sin(\varphi)$ where $\hat{X}$ and $\hat{P}$ represent the amplitude and phase quadratures of the signal mode with $\hat{X} = \frac{\hat{a}_x + \hat{a}_x^+}{2}$ and $\hat{P} = \frac{\hat{a}_x - \hat{a}_x^+}{2i}$.

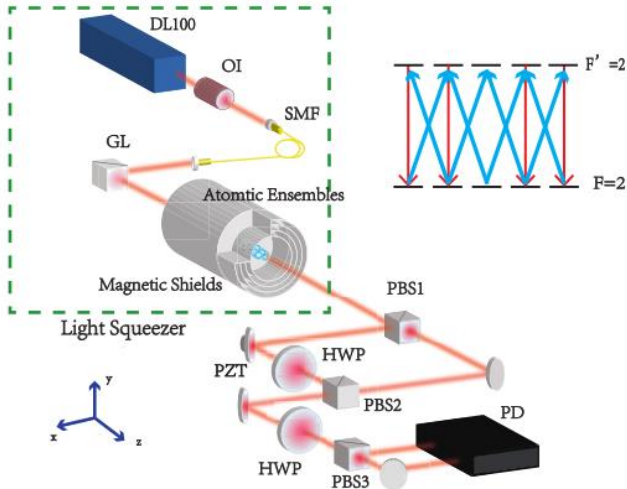

**Fig. 1.** Experiment setup. OI:optical isolator, SMF: single mode fiber, PBS:polarizer beam splitter, GL: Glan polarizer, HWP:half-wave plant, PD:blanced photodetector.

The corresponding standard quantum limit (SQL) of this system is measured by rotating the half-wave plant before the PBS3 carefully to ensure the lowest noise in the spectrum analyzer when the signal field is blocked. In our experiment, different from other experimental researches [6–8, 10, 12, 13] to shield almost all magnetic field around the atomic cells, we apply the weak DC magnetic fields in x- or z- direction (i.e. latitude and longitude direction) inside the magnetic shielding to quantize the atomic spins along x- or z-axis respectively. We use a solenoid to make a pure magnetic field in longitude direction (along x-axis) and use a Helmholtz coil to make a latitude one (along z-axis). The variance of our magnetic field is less then one percent. We know that the magnetic field supplies quantum axe for atoms to quantize atomic spins along its direction. Different quantum axes could construct different interaction configurations by quantizing the collective spins along different quantum axes. In experiment, the polarization of input drive field is fixed along y-direction acting as the optical pumping light and interaction field simultaneously. When the magnetic field is applied in z-direction, the signal mode $E_x$ can be seen as a superposition of $\sigma_+$ and $\sigma_-$ components. In contrast, when the quantization axis along x-direction, the input drive field $E_y$ (seen as $\sigma_+$+ $\sigma_-$) interacts with the atoms to produce the π polarized signal field $E_x$.

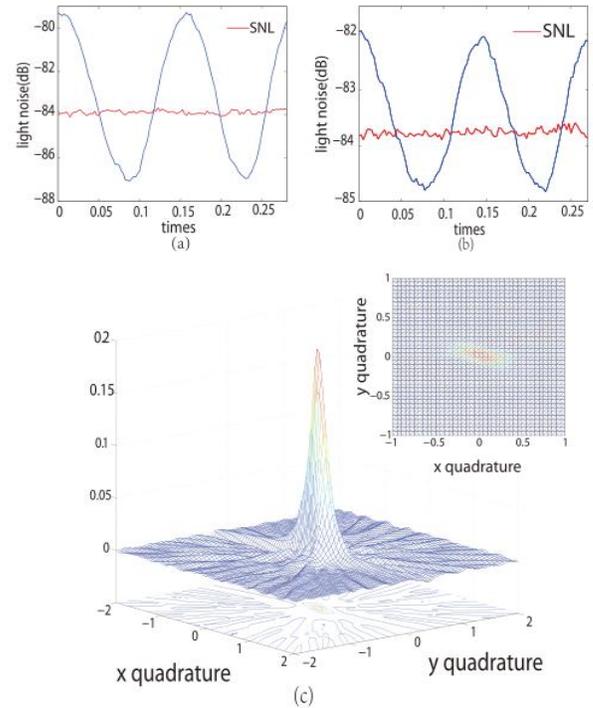

**Fig. 2.** Noise spectrum with (a) and without (b) magnetic field in x-direction. The light power is fixed at 6mW, and laser frequency is locked on resonant to 87Rb D1 transition F = 2 to $F' = 2$ (c)tomography of the vacuum squeezed state, The maximum squeezing occurs near y quadrature.

One noise spectrum without the magnetic field are given in figure 2 (a) when scanning the phase of local oscillator. The noise of rotated light is about 1.0dB below shot noise. Here we focus on the effect of different quantum axis on squeezing. In figure 2 (b), we shows the noise spectrum by applying a magnetic field in x-direction. It is obvious that the squeezing is enhanced 2.0dB. The wigner function of this squeezed state is reducted and the tomography result in figure 2 (c) also shows the squeeze performance. The maximum squeezing occurs nearly quadrature which agrees with the observation in Ref[6].

Figure 3 shows the normalized light noise in shot noise unit as a function of the strength of the applied magnetic fields in two orthogonal directions. The results indicate that increasing the strength of the longitude magnetic field makes the performance of light squeezer deteriorate, which agrees with the observation in Ref[14]. In contrast, strengthening the magnetic field in x-direction can improve the squeeze degree initially, and then it achieves the maximum value and keeps it when the magnetic field is strong enough to maintain the quantization axis of collective spins.

We also characterize the squeeze degree of the squeezed light as function of the frequency and power of driven laser by applying 6mW drive light and 100mG magnetic field in x- direction. The results are shown in Figure 4 (a) and (b). The largest squeezing occurs when the frequency of the driven laser is around F = 2 → F '=2 transition. Squeezing increases with the power of the drive light until reaches the certain level as a balance result of optical pumping and AC Stark shift[10, 15].

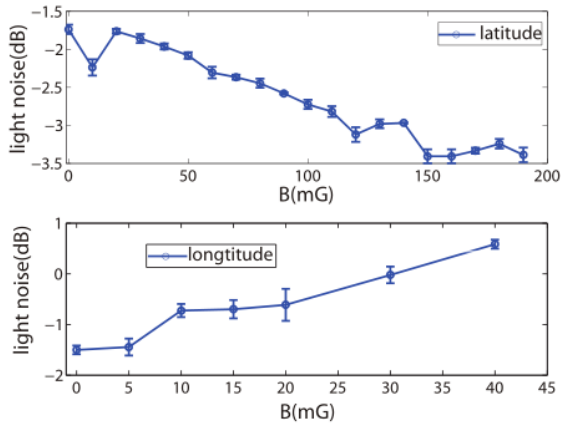

**Fig. 3.** Minimum noise of light quadrature $\hat{X}(\phi)$ normalized to the shot-noise unit as a function of the strength of magnetic field applied in x- (latitude) or z-(longitude) direction. The light power is fixed at 6mW, and laser frequency is locked on resonant to 87Rb D1 transition F = 2 to $F' = 2$.

We also characterize the squeeze degree of the squeezed light as function of the frequency and power of driven laser by applying 6mW drive light and 100mG magnetic field in x-direction. The results are shown in Figure 4 (a) and (b). The largest squeezing occurs when the frequency of the driven laser is around F = 2→F'=2 transition. Squeezing increases with the power of the drive light until reaches the certain level as a balance result of optical pumping and AC Stark shift [10, 15].

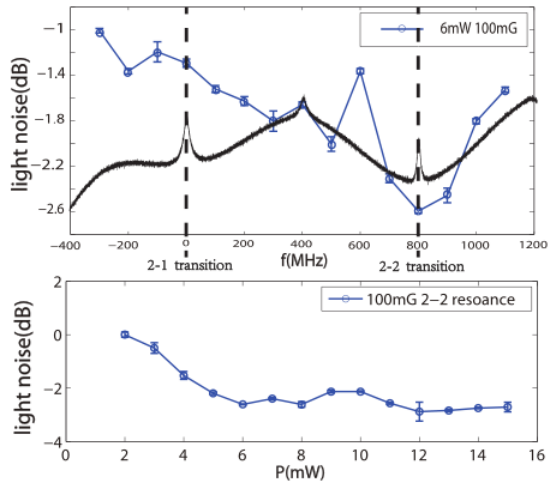

**Fig. 4.** Minimum noise of light quadrature $\hat{X}(\phi)$ normalized to the shot-noise level as a function of detuning (a) and light power (b). In (a) the light power is fixed at 6mW, and a 100mG magnetic field along x-axis is applied. Laser detuning $\triangle$ is defined as $\triangle = \omega_l - \omega_{21}$ where $\omega_{21}$ is the optical frequency corresponding to 87Rb D1 transition F=2 to $F' = 1$ at 795nm. The black line is saturated absorption spectra, characterizing the 87Rb D1 transition. In (b) laser frequency is locked on resonant to 87Rb D1 transition F=2 to $F' = 2$, and a 100mG magnetic field along x-axis is applied.

In conclusion, the maximum squeeze degree in our experiment is -3.5dB by applying 150mG x-direction magnetic field on the atomic vapor and 6mW drive field resonant on F = 2–→ F'=2 transition. The loss of the signal light between the end of the cell and detector is 80% , the quantum efficiency of the detector is 95% and the visibility between signal and local oscillation lights is 99%. The squeeze degree reach -4±0.2dB after considering these effects. In conclusion, we reported a compact squeezed light source at 795 nm based on polarization self-rotation in a hot Rubidium vapor. Compared with other experimental studies, we applied a weak magnetic field on the atomic vapor in x- and z-direction, atoms are oriented in two orthogonal axes, corresponding to different optical pumping and interaction configuration. By optimizing of the frequency and intensity of driven field, we observed -3.5±0.2dB vacuum squeezing below shot noise when applying the magnetic field along x-axis. Since the squeezed light is generated in the atomic ensemble, this compact nonclassical light source can be directly applied in the quantum protocols where atomic ensembles are involved such as quantum repeater and atomic magnetometer.

## FUNDING INFORMATION

**Funding.** National Basic Research Program of China (973 Program grant no.2011CB921604), the National Natural Sci- ence Foundation of China (grant numbers 11274118, 91536114, 11129402 and 11234003),the Program of Indtroducing Talents of Discipline to Universities under Grant B12024 and Supported by Innovation Program of Shanghai Municipal Education Commission (grant no. 13ZZ036).

**Acknowledgment**. We thank Changde Xie and Xiaojun Jia from Shanxi University , Z.Y. Ou from Indiana University-Purdue University Indianapolis for the helpful discussion.